%% file: main.tex
\documentclass[sigconf]{acmart}

\usepackage{algorithm}
\usepackage{multirow}
\usepackage{enumitem}
\usepackage{bm}
\usepackage{xspace}
\usepackage{marvosym}

\usepackage{bbding}
\AtBeginDocument{%
  }

\copyrightyear{2023}
\acmYear{2023}
\setcopyright{acmlicensed}\acmConference[WWW '23]{Proceedings of the ACM
Web Conference 2023}{May 1--5, 2023}{Austin, TX, USA}
\acmBooktitle{Proceedings of the ACM Web Conference 2023 (WWW '23), May
1--5, 2023, Austin, TX, USA}
\acmPrice{15.00}
\acmDOI{10.1145/3543507.3583434}
\acmISBN{978-1-4503-9416-1/23/04}

\begin{document}

\title[Learning Vector-Quantized Item Representation for Transferable Sequential Recommenders]{\texorpdfstring{Learning Vector-Quantized Item Representation for \\Transferable Sequential Recommenders}{Learning Vector-Quantized Item Representation for Transferable Sequential Recommenders}}

\author{Yupeng Hou}
\email{houyupeng@ruc.edu.cn}
\affiliation{
    \institution{Gaoling School of Artificial Intelligence}
    \institution{Renmin University of China}
    \city{Beijing}
    \country{China}
}

\author{Zhankui He}
\email{zhh004@eng.ucsd.edu}
\affiliation{
    \institution{UC San Diego}
    \city{San Diego}
    \state{California}
    \country{USA}
}

\author{Julian McAuley}
\email{jmcauley@eng.ucsd.edu}
\affiliation{
    \institution{UC San Diego}
    \city{San Diego}
    \state{California}
    \country{USA}
}

\author{Wayne Xin Zhao$^{\dagger}$\textsuperscript{\Letter}}
\email{batmanfly@gmail.com}
\affiliation{
    \institution{Gaoling School of Artificial Intelligence}
    \institution{Renmin University of China}
    \city{Beijing}
    \country{China}
}
\thanks{$\dagger$ Beijing Key Laboratory of Big Data Management and Analysis Methods.}
\thanks{\Letter\ Corresponding author.}

\renewcommand{\authors}{Yupeng Hou, Zhankui He, Julian McAuley, Wayne Xin Zhao}
\renewcommand{\shortauthors}{Hou, et al.}

\newcommand{\ie}{\emph{i.e.,}\xspace}
\newcommand{\eg}{\emph{e.g.,}\xspace}
\newcommand{\aka}{\emph{a.k.a.,}\xspace}
\newcommand{\etal}{\emph{et al.}\xspace}
\newcommand{\paratitle}[1]{\vspace{1.5ex}\noindent\textbf{#1}}
\newcommand{\wrt}{w.r.t.\xspace}
\newcommand{\ignore}[1]{}

\newcommand{\tba}{\textcolor{red}{xxx }}
\newcommand{\outd}{\textcolor{red}{[Outdated]}~}
\newcommand{\tabincell}[2]{\begin{tabular}{@{}#1@{}}#2\end{tabular}}

\definecolor{dark2green}{rgb}{0.1, 0.65, 0.3}
\definecolor{dark2orange}{rgb}{0.9, 0.4, 0.}
\definecolor{dark2purple}{rgb}{0.4, 0.4, 0.8}
\newcommand{\first}[1]{\textbf{#1}}
\newcommand{\second}[1]{\underline{#1}}
\newcommand{\third}[1]{\textbf{\textcolor{dark2purple}{#1}}}

\begin{abstract}
Recently, the generality of natural language text has been leveraged to develop transferable recommender systems. The basic idea is to employ pre-trained language models~(PLM) to encode item text
into item representations. Despite the promising transferability, the binding between item text and item representations might be \emph{too tight},
leading to potential problems such as over-emphasizing the effect of text features and exaggerating the negative impact of domain gap. To address this issue, this paper proposes \textbf{VQ-Rec}, a novel approach to learning \underline{V}ector-\underline{Q}uantized item representations for transferable sequential \underline{Rec}ommenders. The main novelty of our approach lies in the new item  representation scheme: it first maps item text into a vector of discrete indices (called \emph{item code}), and then employs these indices to lookup the code embedding table for deriving item representations. Such a scheme can be denoted as ``\emph{text} $\Longrightarrow$ \emph{code} $\Longrightarrow$  \emph{representation}''. %
Based on this representation scheme, we further propose an enhanced contrastive pre-training approach, using semi-synthetic and mixed-domain code representations as hard negatives. Furthermore, we design a new cross-domain fine-tuning method based on a differentiable permutation-based network. 
Extensive experiments conducted on six public benchmarks demonstrate the effectiveness of the proposed approach, in both cross-domain and cross-platform settings.  
Code and pre-trained model are available at: \textcolor{blue}{\url{https://github.com/RUCAIBox/VQ-Rec}}.

\end{abstract}


\maketitle

\input{sec-intro}

\input{sec-method}
\input{sec-experiments}
\input{sec-related}

\section{Conclusions}

In this work, we proposed VQ-Rec to learn vector-quantized item representations for transferable sequential Recommenders. Different from existing approaches that directly map  text encodings from PLMs into item representations, we established a two-step item representation scheme, in which it firstly maps text encodings into  discrete codes and then employs embedding lookup to derive item representations. 
To pre-train our approach on multi-domain interaction data, we employed  both mixed-domain and semi-synthetic code representations as hard negatives. 
We further proposed a permutation-based network to 
learn domain-specific code-embedding alignment, which can effectively adapt to downstream domains.  Extensive experiments conducted on six transferring benchmarks demonstrated the effectiveness of VQ-Rec.

\begin{acks}
This work was partially supported by National Natural Science Foundation of China under Grant No. 62222215,
Beijing Natural Science Foundation under Grant No. 4222027, and  Beijing Outstanding Young Scientist Program under Grant No. BJJWZYJH012019100020098.
Xin Zhao is the corresponding author.
\end{acks}

\bibliographystyle{ACM-Reference-Format}
\balance
\bibliography{main}

\end{document}

%% file: sec-intro.tex
\section{Introduction}

Sequential recommender systems have been widely deployed on various application platforms for recommending 
items of interest
to users. Typically, such a recommendation task is formulated as a sequence prediction problem~\cite{rendle2010fpmc,hidasi2016gru4rec,kang2018sasrec,sun2019bert4rec}, inferring the next item(s) that a user is likely to interact with based on her/his historical interaction sequences. Although a similar task formulation has been taken for different sequential recommenders, 
it is
difficult to reuse an existing well-trained recommender for 
new recommendation scenarios~\cite{li2022recguru,hou2022unisrec}. For example, when a new domain emerges with specific interaction characteristics, one may need to train a recommender  from scratch, which is time-consuming and 
can
suffer from 
cold-start issues. Thus, it is desirable to develop \emph{transferable sequential recommenders}~\cite{ding2021zero,hou2022unisrec,wang2022transrec}, which can quickly adapt to new domains or scenarios. 

For this purpose, in recommender systems literature,  early studies mainly conduct cross-domain recommendation methods~\cite{zhu2019dtcdr,zhu2021crossdomain,li2022recguru} by transferring the learned knowledge from existing domains to a new one. These studies mainly assume that shared information (\eg overlapping users/items~\cite{singh2008cmf,hu2018conet,zhu2019dtcdr} or common features~\cite{tang2012cdcr}) are available for learning cross-domain mapping relations. However, in real applications, users or items are only partially shared or completely non-overlapping across different domains (especially in a cross-platform setting), making it difficult to effectively conduct cross-domain transfer. Besides, previous content-based transfer methods~\cite{tang2012cdcr,fernandez2014exploiting} usually design specific approaches tailored for the data format of shared features, which is not generally useful in various recommendation scenarios.  

As a recent approach, several studies~\cite{ding2021zero,hou2022unisrec,wang2022transrec} propose to leverage the generality of natural language texts (\ie title and description text of items, called \emph{item text}) for bridging the domain gap in recommender systems. 
The basic idea is to employ the learned text encodings via pre-trained language models (PLM)~\cite{sun2019bert4rec,brown2020gpt3} as \emph{universal item representations}. 
Based on such item representations, sequential recommenders pre-trained on the interaction data from a mixture of multiple domains~\cite{ding2021zero,hou2022unisrec,wang2022transrec} have shown promising transferability.
Such a paradigm can be denoted as ``\emph{text} $\Longrightarrow$ \emph{representation}''.
Despite the effectiveness, %
 we argue that the binding between item text and item representations is ``\emph{too tight}'' in previous approaches~\cite{ding2021zero,hou2022unisrec}, thus leading to two potential issues. 
First, since these methods employ text encodings to derive item representations (without using item IDs), text semantics have a direct influence on the recommendation model. Thus, the recommender might over-emphasize the effect of text features (\eg generating very similar recommendations in texts) instead of sequential characteristics reflected in interaction data. Secondly, text encodings from different domains (with varied distributions and semantics~\cite{fan2022adaranker,hou2022unisrec}) are not naturally aligned in a unified semantic space, and 
the domain gap existing in text encodings is likely to cause a performance drop during multi-domain pre-training.  
The tight binding between text encodings and item representations might exaggerate the negative impact of the domain gap.

Considering these issues, our solution is to incorporate  intermediate discrete item indices (called \emph{codes} in this work) in item representation scheme and relax the strong binding between item text and item representations, which can be denoted as  
``\emph{text} $\Longrightarrow$ \emph{code} $\Longrightarrow$  \emph{representation}''.
Instead of directly mapping text encodings into item representations, we consider a two-step item representation scheme. Given an item, it first maps the item text to a vector of discrete indices (\ie item code), and then aggregates the corresponding embeddings according to the item code as the item representation.
The merits of such a representation scheme are twofold. Firstly, item text is mainly utilized to generate discrete codes, which can reduce its influence on the recommendation model meanwhile inject useful text semantics. Second, the two mapping steps can be learned or tuned according to downstream domains or tasks, making it more flexible to fit new recommendation scenarios. 
To develop our approach, we highlight two key challenges to address: (i) how to learn discrete item  codes that are sufficiently distinguishable for accurate  recommendation; (ii) how to effectively pre-train and adapt the item representations considering the varied distribution and semantics across different domains.%

To this end,   we propose \textbf{VQ-Rec}, a novel approach to learn \underline{V}ector-\underline{Q}uantized item representations for transferable sequential \underline{Rec}ommenders.  Different from existing transferable recommenders based on PLM encoding, VQ-Rec maps each item into a discrete $D$-dimensional code as the indices for embedding lookup. %
To obtain semantically-rich and distinguishable item codes, we utilize optimized product quantization (OPQ) techniques to discretize text encodings of items. In this way, \textcolor{black}{the discrete codes} that preserve the textual semantics are  distributed over the item set in a more uniform way, so as to be highly distinguishable.  
Since our representation scheme does not modify the underlying backbone (\ie Transformer), it is generally applicable to various sequential architectures.  
To capture transferable patterns based on item codes, we pre-train the recommender on a mixture of multiple domains in a contrastive learning approach. Both mixed-domain and semi-synthetic code representations are used as hard negatives to enhance the contrastive training. To transfer the pre-trained model to a downstream domain, we propose a differentiable permutation-based network to learn the code-embedding alignment, and further update the code embedding table to fit the new domain. 
Such fine-tuning is highly parameter-efficient, as only the parameters involved in item representations need to be tuned.

Empirically, we conduct extensive experiments on six benchmarks, including both cross-domain and cross-platform scenarios. Experimental results demonstrate the strong transferability of our approach. 
Especially, inductive recommenders purely based on item text can recommend new items without re-training, and meanwhile gain better performance on known items.

%% file: sec-method.tex
\section{Methodology}

In this section, we present the proposed transferable sequential \textbf{Rec}ommendation approach based on \textbf{V}ector-\textbf{Q}uantized item indices, named \textbf{VQ-Rec}. 

\begin{figure*}[ht]
    \centering
    \includegraphics[width=0.85\textwidth]{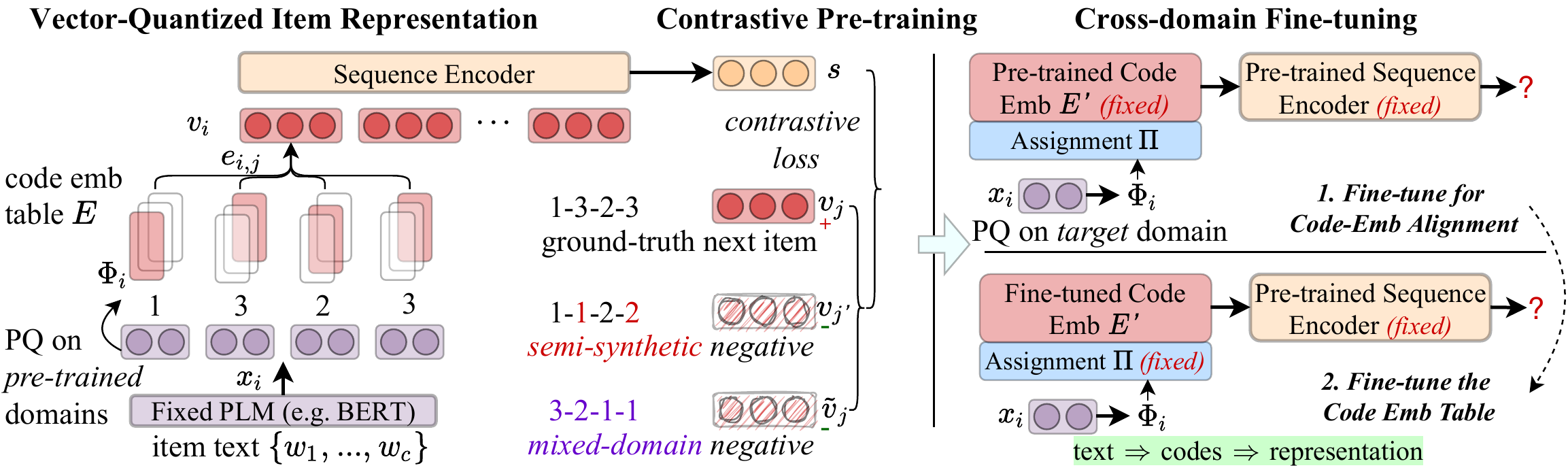}
  \caption{The overall framework of the proposed vector-quantized code-based transferable sequential recommender VQ-Rec.}
    \label{fig:overall}
\end{figure*}

\subsection{Approach Overview}

\paratitle{Task formulation.} We consider the sequential recommendation task setting that \emph{multi-domain}  interaction data is available as training (or pre-training) data. Formally, the interaction data of a user in some domain can be denoted as an interaction sequence $s=\{i_1,i_2,\ldots,i_n\}$ (in chronological order), where each interacted item $i$ is associated with a unique item ID and  text data, \eg title or description (\emph{item text}). Since a user is likely to  interact with items from multiple domains, we can derive multiple interaction sequences for a user. Considering the large semantic gap across different domains~\cite{hou2022unisrec}, we don't combine the multiple interaction sequences of a user into a single sequence, but instead keep these sequences per domain.
Note that the item IDs are not explicitly utilized to generate  item representations in our approach.
The task goal is to pre-train a transferable sequential recommender that can effectively adapt to new domains (unseen in training data).

\paratitle{Solution overview.} To develop the sequential recommender, we adopt the popular Transformer architecture~\cite{kang2018sasrec} as the backbone of our approach. It is built on the self-attention mechanism, taking as input item embeddings and positional embeddings at each time step. Unlike previous related studies~\cite{hou2022unisrec}, we don't include any additional components (\eg adaptors) into the Transformer architecture, but instead learn transferable item representations for feeding the backbone. 
The key novelty of our approach lies in the new item  representation scheme for sequential recommenders.
In this scheme, we first map item text into a vector of discrete indices (called an \emph{item code}), and then employ these indices to lookup the \emph{code embedding table} for deriving item representations. Such a scheme can be denoted as ``\emph{text} $\Longrightarrow$ \emph{code} $\Longrightarrow$  \emph{representation}'', which removes the tight binding between item text and item representations.
In order to learn and transfer such item representations, we further propose specific strategies for \emph{contrastive recommender pre-training} and \emph{cross-domain recommender fine-tuning}. 

The overall framework of the proposed approach VQ-Rec is depicted in Figure~\ref{fig:overall}. We consider three key components for developing transferable recommenders: (i) how to represent the items with vector-quantized code representation (Section~\ref{sec:discrete_code}); (ii) how to train the recommenders based on the new representation scheme  (Section~\ref{sec:pre_train}); (iii) how to transfer the pre-trained recommender to new domains (Section~\ref{sec:transfer}). 

\subsection{Vector-Quantized Item Representation} \label{sec:discrete_code}

As introduced before, we propose a two-step item representation scheme: 
(i) item text is encoded by PLMs into a vector of  discrete codes (Section~\ref{sec:code_learning}), and (ii) discrete codes are employed to lookup the code embedding table to generate the item representation (Section~\ref{sec:item_rep_learning}). Next, we present the representation scheme in detail.

\subsubsection{Vector-Quantized Code Learning} \label{sec:code_learning}
In this part, we first study how to map item text into a discrete code. %
In order to leverage the generality of natural language text, we first encode the descriptive text of items into text encodings via PLMs. Then, we  build a mapping between text encodings and discrete codes based on optimized product quantization. Such a process is described as follows:

\textbf{(1) item text $\stackrel{PLM}{\Longrightarrow}$ text encodings.} We utilize the widely used BERT model~\cite{devlin2019bert} to encode the text information  of items. Given an item $i$, we insert a special token \texttt{[CLS]} at the beginning of its item text $t_i$ (denoted by $\{w_1, \ldots, w_c\}$), and then feed this extended text to BERT 
for obtaining the text encoding of item $i$ as:
\begin{equation}
    \bm{x}_i = \text{BERT}(\left[\texttt{[CLS]};w_1;\ldots;w_c\right]),\label{eqn:bert}
\end{equation}
where $\bm{x}_i \in \mathbb{R}^{d_W}$ is the final hidden vector corresponding to the first input token (\texttt{[CLS]}), and ``[;]'' is the concatenation operation. Note that  the BERT encoder is fixed in our approach. 

\textbf{(2) text encodings $\stackrel{PQ}{\Longrightarrow}$ discrete codes.} Next, we consider mapping the text encoding $\bm{x}_i$ to a vector of discrete codes for item $i$, based on product quantization (PQ)~\cite{jegou2010pq}. PQ defines $D$ sets of vectors, each of which corresponds to $M$ centroid embeddings with dimension $d_W/D$. Formally, we denote $\bm{a}_{k,j}\in\mathbb{R}^{{d_W}/{D}}$ as the $j$-th centroid embedding for the $k$-th vector set. Given the text encoding $\bm{x}_i$, PQ first splits it into $D$ sub-vectors $\bm{x}_i = \left[\bm{x}_{i,1};\ldots;\bm{x}_{i,D}\right]$.
Then for each sub-vector, PQ selects the \emph{index} of the nearest centroid embedding from the corresponding set to compose  its discrete code (\ie a vector of indices). For the $k$-th sub-vector of item $i$, formally we have the selected index as:
\begin{equation}
    c_{i,k} = \arg \min_{j} \|\bm{x}_{i,k}-\bm{a}_{k,j}\|^2 \in \{1,2,\ldots,M\},\label{eqn:index}
\end{equation}
where $c_{i,k}$ is the $k$-th dimension of the discrete code representation for item $i$. To learn these PQ centroid embeddings (\ie $\{\bm{a}_{k,j}\}$), we adopt a popular method \emph{optimized product quantization (OPQ)}~\cite{ge2013opq}, and then optimize  centroid embeddings based on the text encodings of items from all the training domains. After centroid embedding learning, we can perform the index assignment (Eq.~\eqref{eqn:index}) independently for each dimension, and obtain the discrete code representation $\bm{c}_i = \left(c_{i,1},\ldots, c_{i,D} \right)$ for item $i$.

\subsubsection{Code Embedding Lookup as Item Representations}  \label{sec:item_rep_learning}

Given the learned discrete item codes, we can directly perform the embedding lookup to generate item representations. 

Since  our approach uses $D$-dimensional discrete indices as item codes, denoted by $\left(c_{i,1},\ldots, c_{i,D} \right)$, we set up $D$ \emph{code embedding matrices} (called \emph{code embedding table}) for lookup, each of which corresponds a dimension of the code representation. 
For each dimension $k$, the discrete codes of all the items share a common code embedding matrix $\bm{E}^{(k)} \in \mathbb{R}^{M \times d_V}$, where $d_V$ is the dimension of the final item representations. Next we can obtain the code embeddings for item $i$ by embedding lookup as $\{\bm{e}_{1,c_{i,1}},\ldots,\bm{e}_{D,c_{i,D}}\}$, where $\bm{e}_{k,c_{i,k}}\in \mathbb{R}^{d_V}$ is the $c_{i,k}$-th row of parameter matrix $\bm{E}^{(k)} $. 
Note that 
the code embedding $\bm{e}_{k,j} \in \mathbb{R}^{d_V}$ is different from the corresponding PQ centroid embedding $\bm{a}_{k,j} \in \mathbb{R}^{d_W/D}$.
We randomly initialize $\bm{E}_k$ for pre-training.
After getting the code embeddings of item $i$, we further aggregate them to generate
the final item representation:%
\begin{equation}
    \bm{v}_i = \operatorname{Pool}\left(\left[\bm{e}_{1,c_{i,1}};\ldots;\bm{e}_{D,c_{i,D}}\right]\right),\label{eqn:item_rep}
\end{equation}
where $\bm{v}_i \in \mathbb{R}^{d_V}$ is the derived item representations, and $\operatorname{Pool}(\cdot): \mathbb{R}^{D \times d_V}\to\mathbb{R}^{d_V}$ is the mean pooling function. %

\subsubsection{Representation Distinguishability \emph{vs.}~Code Uniformity} \label{sec:distinguish}
To make accurate recommendations, the item representations should be distinguishable over a large candidate space, especially for those with similar text encodings. In our case, we should try to avoid the collision (\ie assigning the same code for two items) in the discrete codes between any two items.
It has been shown by \citet{zhan2022repconc} that the minimum collision probability is achieved when vectors are equally quantized to all possible discrete codes.
As a result, ideally, the discrete codes should be \emph{uniformly} distributed for maximum distinguishability, \ie code $c_{i,k}$ has equal probability of be each of $\{1,2,\ldots,M\}$. 
According to previous empirical results~\cite{xiong2006kmeans_uniformty,zhan2021jpq,zhan2022repconc},
the used technique for training OPQ, \ie $K$-means, tends to generate clusters with a relatively uniform distribution on the cluster sizes. Such results indicate that OPQ can generate discrete code representations with strong distinguishability for items.

\subsection{Contrastive Recommender Pre-training%
} \label{sec:pre_train}

Compared with direct text mapping methods~\cite{ding2021zero,hou2022unisrec}, it is more difficult to optimize our item representation approach, since it involves discrete codes and employs a two-step representation mapping.  
In the following, we first introduce the sequential encoder architecture and then present our proposed contrastive pre-training tasks.
To improve the sequential recommenders training, we propose to use both \emph{mixed-domain} and \emph{semi-synthetic} negative instances.

\subsubsection{Self-attentive Sequence Encoding} Given a sequence of item representations  derived from the vector-quantized item codes, we use a sequential encoder to obtain the sequence representation. Following SASRec~\cite{kang2018sasrec}, we adopt a widely-used self-attentive Transformer architecture~\cite{vaswani2017attention}.
In detail, a typical Transformer encoder consists of stacks of multi-head self-attention layers (denoted by $\operatorname{Attn}(\cdot)$) and multilayer perceptron networks (denoted by $\operatorname{FFN}(\cdot)$). The index representations $\bm{v}_i$ (Eqn.~\eqref{eqn:item_rep}) and absolute position embeddings $\bm{p}_j$ are summed up as input of Transformer encoder at position $j$. Then the update process can be formally written as:
\begin{align}
    \bm{f}_j^0 &= \bm{v}_i + \bm{p}_j, \\
    \bm{F}^{l+1} &= \operatorname{FFN}\left(\operatorname{Attn}\left(\bm{F}^l\right)\right), l\in\{1,2,\ldots,L\},\label{eqn:seq_rep}
\end{align}
where $\bm{F}^l = [\bm{f}_0^l;\ldots;\bm{f}_n^l] \in \mathbb{R}^{n\times d_V}$ denotes the hidden states at each  position in the $l$-th layer. We take the final hidden state $\bm{f}_n^L$ corresponding to the $n$-th (last) position as the \emph{sequence representation}.

\subsubsection{Enhanced Contrastive Pre-training}

To optimize the sequential recommender, a commonly used approach is to conduct a batch-level optimization objective with contrastive learning~\cite{chen2020simclr,xie2022cl4rec,lin2022ncl}.
Specifically, for a batch of $B$ training instances, where each instance is a pair of the sequential context (\ie historical interaction sequence) and the ground-truth next item (\ie positives). 
We encode them into representations $\{\langle\bm{s}_1, \bm{v}_1\rangle, \ldots,\langle\bm{s}_B, \bm{v}_B\rangle\}$, where $\bm{s}_{j}$ denotes the $j$-th normalized sequence representations (Eqn.~\eqref{eqn:seq_rep}) and $\bm{v}_{j}$ denotes the $j$-th normalized  representation of positive item paired with $\bm{s}_{j}$ (Eqn.~\eqref{eqn:item_rep}). 
For conducting  contrastive training, a key step is to sample a number of negatives to contrast the positive, which are usually obtained by random sampling. 

However, random negative sampling 
cannot
work well in our approach, due to two major reasons. 
Firstly, since our representation scheme involves the discrete code, it generates an exponential-sized combination of discrete indices (size $M^D$ for $D$ sets and $M$ vectors in each set). While, the set of observed discrete indices in training data  will be much smaller, which causes the \emph{representation sparsity} issue. Besides, we also need to effectively alleviate the domain gap when learning with multi-domain training data. 
Considering the two issues, we design two types of enhanced negatives accordingly.

\paratitle{Semi-synthetic negatives.} Besides the item indices that exist in the training set, we also consider synthesizing augmented item indices as negatives to alleviate the representation sparsity issue. %
However, fully-synthesized indices may be far away from ground-truth items in the sparse code representation space, which will be uninformative to guide  contrastive learning~\cite{kalantidis2020hard}. 
 As a result, we consider generating semi-synthetic codes based on true item codes as \emph{hard negatives}. Given a true item code $\bm{c}_i$, the idea is to randomly replace each index according to a Bernoulli probability parameterized by $\rho \in (0, 1)$, while keeping the remaining indices unchanged. %
 In this way, the item representations derived from semi-synthetic code can be given as follows:
\begin{equation}
    \widetilde{\bm{v}}_i = \operatorname{Emb-Pool}\left(\operatorname{G}(\bm{c}_i)\right),\label{eq:embpool}
\end{equation}
where $\widetilde{\bm{v}}_i$ is the representation of a semi-synthetic hard negative instance, $\operatorname{Emb-Pool}(\cdot)$ is the embedding lookup and aggregation operations described in Section~\ref{sec:item_rep_learning}, and $\operatorname{G(\cdot)}$ is a point-wise generation function:
\begin{equation}
\operatorname{G}(c_{i,j}) = \left\{
    \begin{aligned}
        \ \operatorname{Uni}(\{1,  \ldots, M\}) &, & X=1 \\
        \ c_{i,j} &, & X=0
    \end{aligned}
\right.,
\end{equation}
where $X\sim\operatorname{Bernoulli(\rho)}$, and $\operatorname{Uni}(\cdot)$ samples the item code from the input set uniformly.
Note that  uniform sampling ensures the codes from the semi-synthetic indices follow a similar distribution as true items, which are shown to be distinguishable in Section~\ref{sec:item_rep_learning}.

\paratitle{Mixed-domain negatives.} 
Different from previous next-item prediction models that use in-domain negatives~\cite{hidasi2016gru4rec,kang2018sasrec}, we adopt mixed-domain items as negatives for enhancing the multi-domain fusion during pre-training. Due to efficiency concerns, we directly use the \emph{in-batch} items as negatives (\ie the $B-1$ ground-truth items paired with other sequential contexts).
Since we construct the batch by sampling from multiple domains,
in-batch sampling can naturally generate mixed-domain negatives. 

By integrating the two kinds of negatives, the pre-training objective can be formulated as:
\begin{equation}
    \ell = -\frac{1}{B}\sum_{j=1}^{B} \log \frac{\exp{\left(\bm{s}_j\cdot\bm{v}_j/\tau\right)}}{\underbrace{\exp{\left(\bm{s}_j\cdot\widetilde{\bm{v}}_{j}/\tau\right) }}_{\text{semi-synthetic}} + \sum_{j'=1}^{B} \underbrace{\exp{\left(\bm{s}_j\cdot\bm{v}_{j'}/\tau\right)}}_{\text{mixed-domain}}},\label{eqn:loss}
\end{equation}
where $\tau$ is a temperature hyper-parameter. Note that here we include the positive in the mixed-domain samples for simplifying the notation, \ie  one positive and $B-1$ negatives.

\subsection{Cross-domain Recommender Fine-tuning} \label{sec:transfer}

In the pre-training stage, we optimize the parameters in the code embedding matrices and Transformer, while the BERT encoder is fixed and PQ is performed independent of pre-training.  Next, we continue to discuss how to conduct the fine-tuning in a cross-domain or cross-platform setting. During fine-tuning, we fix the Transformer sequence encoder (transferred across different domains), and only optimize the parameters involved in item representations, enabling  parameter-efficient fine-tuning. 

To further leverage the learned knowledge from pre-trained recommenders, we consider learning to transfer the code representation scheme ($M\times D$ PQ indices) and the code embedding table ($M$ embedding matrices). Considering the two aspects, we decompose the fine-tuning optimization into two stages, namely fine-tuning code-embedding alignment and the code embedding table.

\subsubsection{Fine-tuning for Code-Embedding Alignment} \label{sec:emb_assign}

In order to transfer the item representation scheme, a straightforward approach is to directly reuse the discrete index set and corresponding embedding table.
However, such a simple way neglects the large semantic gap across different domains, thus leading to a weak transfer capacity to downstream domains. Our solution is to only reuse the discrete index set, and rebuild the mapping from indices to embeddings.

\paratitle{Permutation-based code-embedding alignment.}
To generate the item codes in a downstream domain, we re-train new PQ centroids for capturing domain-specific semantic characteristics. By sharing the code set, we map a new item $i$ into $\bm{\hat{c}}_{i} \in \mathbb{N}_{+}^{D}$ through Eqn.~\eqref{eqn:bert}-\eqref{eqn:index}, with the same number of sets $D$ and sub-vectors $M$.  Since we keep all the discrete indices and code embeddings, we employ a permutation-based approach to re-learning the mapping relations (\ie new lookup scheme) to associate indices with code embeddings. 
For each dimension $k$ of the discrete indices, we formulate the embedding alignment as a matrix $\bm{\Pi}_k \in \{0,1\}^{M\times M}$.
To enforce a bijection alignment, $\bm{\Pi}_k$ should be a permutation matrix that has exactly one entry of $1$ in each row and each column and $0$s elsewhere.
Formally, the aligned code embedding table under a new domain $\bm{\hat{E}}^{(k)}$ can be given as:
\begin{equation}\label{ekplus}
    \bm{\hat{E}}^{(k)}= \bm{\Pi}_k \cdot \bm{E}^{(k)}.
\end{equation}

\paratitle{Alignment optimization.} To learn the code-embedding alignment matrices $\bm{\Pi}_k$, we optimize the corresponding parameters by a conventional next-item prediction objective. Given the sequential context, we can predict the next item according to the following probability:
\begin{equation}
    P(i_{t+1}|i_1,\ldots,i_{t}) = \operatorname{Softmax}(\bm{\hat{s}}\cdot\bm{\hat{v}}_{i_{t+1}}),\label{eqn:p_next_item}
\end{equation}
where $\bm{\hat{s}}$ is the output of the pre-trained sequential encoder that takes as input $\bm{\hat{v}}_{i_1}, \ldots, \bm{\hat{v}}_{i_t}$, and the item representations $\bm{\hat{v}}_i$ can be derived based on the code $\bm{\hat{c}}_{i}$ in a similar way in Eqn.~\eqref{eq:embpool}.
To make the permutation matrices $\bm{\Pi}_k$ differentiable, we are inspired by Birkhoff's theorem~\cite{tay2020sparse}: any doubly stochastic matrix can be considered as a convex combination of permutation matrices. Following this idea, we use doubly stochastic matrices to simulate the permutation. Concretely, we start by randomly initializing a parameter matrix for each dimension $\bm{\Theta}_k \in \mathbb{R}^{M\times M}$, then  convert it into  the doubly stochastic matrix via Gumbel-Sinkhorn algorithm~\cite{mena2018gumbel_sinkhorn}.
Then, we optimize $\{\bm{\Theta}_1, \ldots, \bm{\Theta}_D\}$ according to the next item probability (Eqn.~\eqref{eqn:p_next_item}) using cross-entropy loss, while fixing  the remaining parameters in the pre-trained recommender.

\subsubsection{Fine-tuning the Code Embedding Table}
After code-embedding alignment, we continue to fine-tune the permuted code embedding table for increasing the representation capacity of fitting the downstream domains. To be specific, we  optimize the parameters $\bm{\hat{E}}^{(1)}, \ldots, \bm{\hat{E}}^{(D)}$  in Eqn.~\eqref{ekplus} according to the next-item prediction loss in Eqn.~\eqref{eqn:p_next_item}.
In this stage, the parameters $\{\bm{\Theta}_1, \ldots, \bm{\Theta}_D\}$ in Section~\ref{sec:emb_assign} are fixed. 
The fine-tuned VQ-Rec does not rely on item IDs and can be applied in an \emph{inductive setting}, \ie recommending  new items without re-training the model. When a new item emerges, one can encode its item text into discrete indices, and then obtain the corresponding item representations by embedding lookup. 

\subsection{Discussion}
In this part, we highlight the merits of the proposed VQ-Rec approach in the following three aspects.

$\bullet$~\emph{Capacity}. By leveraging the generality of text semantics, the learned discrete codes and code embeddings can effectively capture transferable patterns across different domains. Unlike existing related studies~\cite{ding2021zero,hou2022unisrec},
our approach doesn't directly map the text encodings into item representations, thus it can avoid over-emphasizing the text similarity  and is also more robust to the noise from text data. Besides, as discussed in Section~\ref{sec:distinguish}, our approach can generate highly distinguishable code representations.

$\bullet$~\emph{Flexibility}. As another difference with previous studies, we don't modify the underlying sequence encoder (\ie Transformer), without any minor change such as  the incorporation of adaptors. Besides, text encoder and PQ  are  independent (in terms of optimization) from the underlying sequence encoder. These decoupled designs make it flexible to extend our approach with various choices of PLM, sequential encoder and  discrete coding method.

$\bullet$~\emph{Efficiency}. 
Our approach introduces three specific designs for efficient model training and utilization: (1) fixed text encoder; (2) independently learned discrete codes;  (3) fixing the sequence encoder during fine-tuning. 
Besides, compared to existing methods like UniSRec~\cite{hou2022unisrec}, VQ-Rec has better time complexity when deriving transferable item representations ($O(d_V D)$ \emph{vs.} $O(d_W d_V D)$). As shown in Section~\ref{sec:item_rep_learning}, VQ-Rec does not require any matrix multiplication as previous methods based on adaptors.

%% file: sec-experiments.tex
\begin{table}[!t] %
	\caption{Statistics of the preprocessed datasets. ``Avg.~$n$'' denotes the average length of item sequences. ``Avg.~$c$'' denotes the average number of words in item text.
	}
	\label{tab:dataset}
	\resizebox{\columnwidth}{!}{
	\begin{tabular}{l *{5}{r}}
		\toprule
		\textbf{Datasets} & \textbf{\#Users} & \textbf{\#Items} & \textbf{\#Inters.} & \textbf{Avg. $n$} & \textbf{Avg. $c$}\\
		\midrule
		\textbf{Pre-trained} & 1,361,408 & 446,975 & 14,029,229 & 13.51 & 139.34 \\
		\midrule
		\textbf{Scientific}  &  8,442 &  4,385 &  59,427 & 7.04 & 182.87 \\
		\textbf{Pantry}      & 13,101 &  4,898 & 126,962 & 9.69 & 83.17 \\
		\textbf{Instruments} & 24,962 &  9,964 & 208,926 & 8.37 & 165.18 \\
		\textbf{Arts}        & 45,486 & 21,019 & 395,150 & 8.69 & 155.57 \\
		\textbf{Office}      & 87,436 & 25,986 & 684,837 & 7.84 & 193.22 \\
		\textbf{Online Retail} & 16,520 & 3,469 & 519,906 & 26.90 & 27.80 \\
		\bottomrule
	\end{tabular}
	}
\end{table}

\section{Experiments}

In this section, we empirically demonstrate the effectiveness and transferability of the proposed approach VQ-Rec. 

\begin{table*}[!ht]
\centering
\caption{Performance comparison of different models. The best and the second-best 
performance is
denoted in bold and underlined fonts, respectively. ``R@K'' is short for ``Recall@K'' and ``N@K'' is short for ``NDCG@K'', respectively. The features used for item representations of each compared model have been listed, whether ID, text (T), or both (ID+T).}
\label{tab:exp-main}
\resizebox{2.1\columnwidth}{!}{
\begin{tabular}{@{}cccccccccccc@{}}
\toprule
\textbf{Scenario} & \textbf{Dataset} & \textbf{Metric} & \textbf{SASRec} \textsubscript{ID} & \textbf{BERT4Rec} \textsubscript{ID} & \textbf{FDSA} \textsubscript{ID+T} & \textbf{S$^3$-Rec} \textsubscript{ID+T} & \textbf{RecGURU} \textsubscript{ID} & \textbf{SASRec} \textsubscript{T} & \textbf{ZESRec} \textsubscript{T} & \textbf{UniSRec} \textsubscript{T} & \textbf{VQ-Rec} \textsubscript{T} \\\midrule \midrule
\multirow{20}{*}{\shortstack{Cross-\\Domain}} &
\multirow{4}{*}{Scientific} &
     R@10 & 0.1080 & 0.0488 & 0.0899 & 0.0525 & 0.1023 & 0.0994 & 0.0851 & \second{0.1188} & \first{0.1211} \\
 & & N@10   & 0.0553 & 0.0243 & 0.0580 & 0.0275 & 0.0572 & 0.0561 & 0.0475 & \second{0.0641} & \first{0.0643} \\
 & & R@50 & 0.2042 & 0.1185 & 0.1732 & 0.1418 & 0.2022 & 0.2162 & 0.1746 & \first{0.2394} & \second{0.2369} \\
 & & N@50   & 0.0760 & 0.0393 & 0.0759 & 0.0468 & 0.0786 & 0.0815 & 0.0670 & \first{0.0903} & \second{0.0897} \\
\cmidrule(l){2-12}
 & \multirow{4}{*}{Pantry} &
     R@10 & 0.0501 & 0.0308 & 0.0395 & 0.0444 & 0.0469 & 0.0585 & 0.0454 & \second{0.0636} & \first{0.0660} \\
 & & N@10   & 0.0218 & 0.0152 & 0.0209 & 0.0214 & 0.0209 & 0.0285 & 0.0230 & \first{0.0306} & \second{0.0293} \\
 & & R@50 & 0.1322 & 0.1030 & 0.1151 & 0.1315 & 0.1269 & 0.1647 & 0.1141 & \second{0.1658} & \first{0.1753} \\
 & & N@50   & 0.0394 & 0.0305 & 0.0370 & 0.0400 & 0.0379 & 0.0523 & 0.0378 & \first{0.0527} & \first{0.0527} \\
 \cmidrule(l){2-12}
 & \multirow{4}{*}{Instruments} &
     R@10 & 0.1118 & 0.0813 & 0.1070 & 0.1056 & 0.1113 & 0.1127 & 0.0783 & \second{0.1189} & \first{0.1222} \\
 & & N@10   & 0.0612 & 0.0620 & \first{0.0796} & 0.0713 & 0.0681 & 0.0661 & 0.0497 & 0.0680 & \second{0.0758} \\
 & & R@50 & 0.2106 & 0.1454 & 0.1890 & 0.1927 & 0.2068 & 0.2104 & 0.1387 & \second{0.2255} & \first{0.2343} \\
 & & N@50   & 0.0826 & 0.0756 & \second{0.0972} & 0.0901 & 0.0887 & 0.0873 & 0.0627 & 0.0912 & \first{0.1002} \\
 \cmidrule(l){2-12}
 & \multirow{4}{*}{Arts} &
     R@10 & \second{0.1108} & 0.0722 & 0.1002 & 0.1003 & 0.1084 & 0.0977 & 0.0664 & 0.1066 & \first{0.1189} \\
 & & N@10   & 0.0587 & 0.0479 & \first{0.0714} & 0.0601 & 0.0651 & 0.0562 & 0.0375 & 0.0586 & \second{0.0703} \\
 & & R@50 & 0.2030 & 0.1367 & 0.1779 & 0.1888 & 0.1979 & 0.1916 & 0.1323 & \second{0.2049} & \first{0.2249} \\
 & & N@50   & 0.0788 & 0.0619 & \second{0.0883} & 0.0793 & 0.0845 & 0.0766 & 0.0518 & 0.0799 & \first{0.0935} \\
 \cmidrule(l){2-12}
 & \multirow{4}{*}{Office} &
     R@10 & 0.1056 & 0.0825 & 0.1118 & 0.1030 & \second{0.1145} & 0.0929 & 0.0641 & 0.1013 & \first{0.1236} \\
 & & N@10   & 0.0710 & 0.0634 & \first{0.0868} & 0.0653 & 0.0768 & 0.0582 & 0.0391 & 0.0619 & \second{0.0814} \\
 & & R@50 & 0.1627 & 0.1227 & 0.1665 & 0.1613 & \second{0.1757} & 0.1580 & 0.1113 & 0.1702 & \first{0.1957} \\
 & & N@50   & 0.0835 & 0.0721 & \first{0.0987} & 0.0780 & 0.0901 & 0.0723 & 0.0493 & 0.0769 & \second{0.0972} \\
 \midrule
\multirow{4}{*}{\shortstack{Cross-\\Platform}} &
\multirow{4}{*}{\shortstack{Online\\ Retail}} &
     R@10 & 0.1460 & 0.1349 & \second{0.1490} & 0.1418 & 0.1467 & 0.1380 & 0.1103 & 0.1449 & \first{0.1557} \\
 & & N@10   & 0.0675 & 0.0653 & \second{0.0719} & 0.0654 & 0.0658 & 0.0672 & 0.0535 & 0.0677 & \first{0.0730} \\
 & & R@50 & \second{0.3872} & 0.3540 & 0.3802 & 0.3702 & 0.3885 & 0.3516 & 0.2750 & 0.3604 & \first{0.3994} \\
 & & N@50   & 0.1201 & 0.1131 & \second{0.1223} & 0.1154 & 0.1188 & 0.1137 & 0.0896 & 0.1149 & \first{0.1263} \\
 \bottomrule
\end{tabular}
}
\end{table*}

\subsection{Experimental Setup}

\subsubsection{Datasets}

We conduct experiments on public benchmarks for transferable recommender evaluation.
Five domains from Amazon review datasets~\cite{ni2019amazon} are used for pre-training (\textit{Food}, \textit{Home}, \textit{CDs}, \textit{Kindle}, and \textit{Movies}). Then the pre-trained model will be transferred to five downstream cross-domain datasets (\textit{Scientific}, \textit{Pantry}, \textit{Instruments}, \textit{Arts}, and \textit{Office}, all from Amazon) and one cross-platform dataset (\textit{Online Retail}\footnote{https://www.kaggle.com/carrie1/ecommerce-data}, a UK-based online retail platform).

Following \citet{hou2022unisrec}, we filter users and items with fewer than five interactions. We then group the interactions in each sub-dataset by users and sort them in chronological order. For the 
descriptive item text, we concatenate fields including \textit{title}, \textit{categories}, and \textit{brand} for sub-datasets from Amazon, and use the \textit{Description} field in the Online Retail dataset. The item text is truncated to a length of 512. The statistics of the preprocessed datasets are summarized in Table~\ref{tab:dataset}. 

\begin{table*}[t]
\centering
\caption{Ablation analysis on three downstream datasets.
The best and the second-best performance is denoted in bold and underlined fonts, respectively.}
\label{tab:exp-ablation}
\resizebox{2.1\columnwidth}{!}{
\begin{tabular}{@{}lcccccccccccc@{}}
\toprule
\multirow{2}{*}{\textbf{Variants}} & \multicolumn{4}{c}{\textbf{Scientific}} & \multicolumn{4}{c}{\textbf{Office}} & \multicolumn{4}{c}{\textbf{Online Retail}} \\
\cmidrule(lr){2-5} \cmidrule(lr){6-9} \cmidrule(lr){10-13}
 & R@10 & N@10 & R@50 & N@50 & R@10 & N@10 & R@50 & N@50 & R@10 & N@10 & R@50 & N@50 \\
\midrule
(0) VQ-Rec                        & \first{0.1211} & \first{0.0643} & \first{0.2369} & \first{0.0897} & \second{0.1236} & 0.0814 & \first{0.1957} & \second{0.0972} & \second{0.1557} & \second{0.0730} & \first{0.3994} & \second{0.1263} \\
\cmidrule(lr){1-1}
(1) \ \ $w/o$ Pre-training        & 0.1104 & 0.0593 & 0.2238 & 0.0839 & 0.1108 & 0.0666 & 0.1804 & 0.0818 & 0.1535 & 0.0717 & 0.3953 & 0.1244 \\
(2) \ \ $w/o$ Semi-synthetic NS   & \second{0.1194} & \second{0.0631} & \second{0.2358} & \second{0.0886} & 0.1227 & 0.0809 & 0.1951 & 0.0968 & 0.1529 & 0.0702 & \second{0.3938} & 0.1230 \\
\cmidrule(lr){1-1}
(3) \ \ $w/o$ Fine-tuning         & 0.0640 & 0.0361 & 0.0909 & 0.0421 & 0.0261 & 0.0150 & 0.0373 & 0.0174 & 0.0197 & 0.0095 & 0.0419 & 0.0142 \\
(4) \ \ Reuse PQ Index Set        & 0.1168 & 0.0618 & 0.2267 & 0.0859 & 0.1182 & 0.0773 & 0.1869 & 0.0922 & 0.1521 & 0.0707 & 0.3917 & 0.1232 \\
(5) \ \ $w/o$ Code-Emb Alignment  & 0.1183 & 0.0612 & 0.2355 & 0.0867 & \first{0.1242} & \second{0.0824} & \second{0.1956} & \first{0.0980} & 0.1515 & 0.0695 & 0.3924 & 0.1224 \\
(6) \ \ Random Code               & 0.0905 & 0.0494 & 0.1769 & 0.0683 & 0.1134 & \first{0.0837} & 0.1698 & 0.0960 & \first{0.1589} & \first{0.0769} & 0.3933 & \first{0.1282} \\
\bottomrule
\end{tabular}
}
\end{table*}

\subsubsection{Compared Methods}

We compare the proposed approach with the following baseline methods:

$\bullet$ \textbf{SASRec}~\cite{kang2018sasrec} utilizes a self-attentive model to capture item correlations. We implement two versions, either with (1) conventional ID embeddings, or (2) fine-tuned BERT representations of item text as basic item representations.

$\bullet$ \textbf{BERT4Rec}~\cite{sun2019bert4rec} adopts a bi-directional self-attentive model with a cloze objective for sequence modeling.

$\bullet$ \textbf{FDSA}~\cite{zhang2019fdsa} models item and feature sequences with separate self-attentive sequential models.

$\bullet$ \textbf{S$^3$-Rec}~\cite{zhou2020s3rec} captures feature-item correlations at the pre-training stage with mutual information maximization objectives.

$\bullet$ \textbf{RecGURU}~\cite{li2022recguru} proposes an adversarial learning paradigm to pre-train user representations via an auto-encoder. %

$\bullet$ \textbf{ZESRec}~\cite{ding2021zero} encodes item text with PLM as basic item representations. For fair comparison, ZESRec is pre-trained on the same datasets as VQ-Rec.

$\bullet$ \textbf{UniSRec}~\cite{hou2022unisrec} equips textual item representations with an MoE-enhanced adaptor for domain fusion and adaptation. Both item-sequence and sequence-sequence contrastive learning tasks are designed for pre-training transferable sequence representations.

For our approach, we first pre-train one \textbf{VQ-Rec} model on the mixture of item sequences from the pre-training datasets. The pre-trained model will be fine-tuned to each downstream dataset. 

\subsubsection{Evaluation Settings}

Following previous works~\cite{zhou2020s3rec,hou2022unisrec,zhao2022evaluation}, we adopt two widely used ranking metrics, Recall@K and NDCG@K, where K $\in\{10,50\}$. For dataset splitting, we apply the leave-one-out strategy, \ie the latest interacted item as test data, the item before the last one as validation data. The ground-truth item of each sequence is ranked among all the other items while evaluating. We finally report the average scores of all test users.

\subsubsection{Implementation Details}

We implement our models based on Faiss ANNS library~\cite{johnson2021faiss} and \textsc{RecBole}~\cite{zhao2021recbole,zhao2022recbole2}. We use $(M=32) \times (D=256)$ PQ indices as the code representation scheme. We pre-train VQ-Rec for 300 epochs with temperature $\tau=0.07$ and semi-synthetic ratio $\rho=0.75$. 
The iteration number for Gumbel-Sinkhorn algorithm is set to $3$.
The main baseline results are taken from~\citet{hou2022unisrec} directly. 
For other models, we
search the hyper-parameters to find optimal results. The batch size is set to 2,048. The learning rate is tuned in $\{0.0003, 0.001, 0.003\}$. The number of permutation learning epochs is tuned in $\{3,5,10\}$. The models with the highest NDCG@10 results on the validation set are selected for evaluation on the test set. We adopt early stopping with a patience of 10 epochs.

\subsection{Overall Performance}

We compare VQ-Rec with the baseline methods on six benchmarks and report the results
in Table~\ref{tab:exp-main}.

For the baseline methods, we can see that text-based models (\ie SASRec (T), ZESRec, and UniSRec) perform better than other approaches on small-scale datasets (\eg Scientific and Pantry). For these datasets where interactions are not sufficient enough to train a powerful ID-based recommender, the text-based methods may benefit from the text  characteristics. While for the models that incorporate item IDs (\ie SASRec and BERT4Rec), they perform better on those datasets with more interactions (\eg Arts, Office and Online Retail), showing that over-emphasizing the text similarity may lead to sub-optimal results.

For the proposed approach VQ-Rec, it achieves the best or the second-best performance on all  datasets. On small-scale datasets, the results show that the proposed discrete indices can preserve text semantics to give proper recommendations. On large-scale datasets, VQ-Rec can be well-trained to capture sequential characteristics.
Note that VQ-Rec can be applied in an inductive setting, which can recommend new items without re-training the model. The experimental results also show that with carefully designed encoding schemes as well as large-scale pre-training, inductive recommenders can also outperform conventional transductive models.

\begin{figure}[t]
	{
		\begin{minipage}[t]{0.49\linewidth}
			\centering
			\includegraphics[width=1\textwidth]{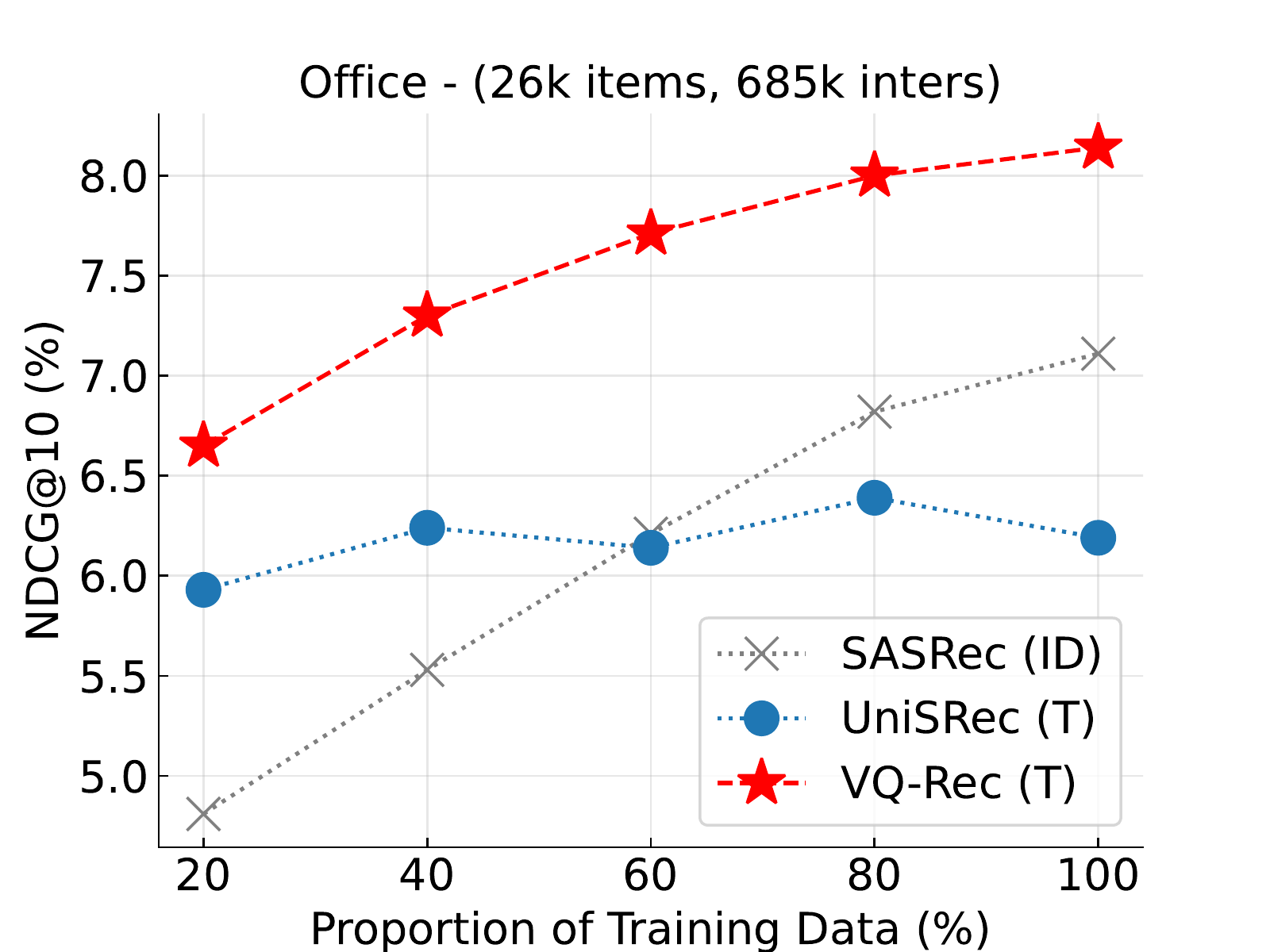}
		\end{minipage}
		\begin{minipage}[t]{0.49\linewidth}
			\centering
			\includegraphics[width=1\textwidth]{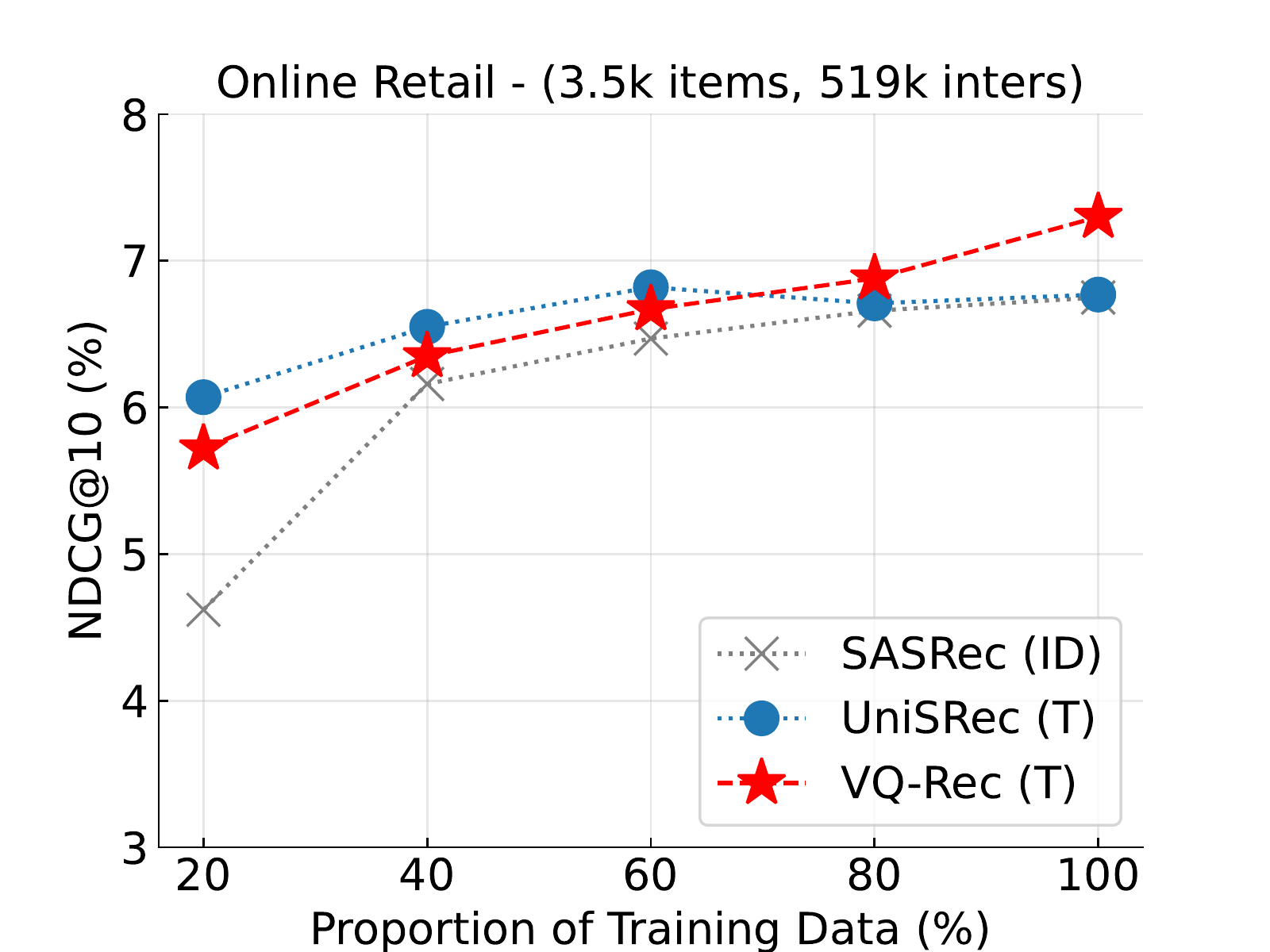}
		\end{minipage}
	}
	\caption{Performance comparison \wrt increasing training data on ``Office'' and ``Online Retail'' datasets. The x-axis coordinates denote the proportion of original training data that the recommenders are trained on.}
	\label{fig:exp-sparse}
\end{figure}

\subsection{Ablation Study}

We analyze how each of the proposed components affects final performance. Table~\ref{tab:exp-ablation} shows the performance of our default method and its six variants on three representative datasets, including one small-scale dataset (Scientific), one large-scale dataset (Office), and one cross-platform dataset (Online Retail).

$(1)$ \emph{$w/o$ Pre-training}: Without pre-training on multiple domains, the variant performs worse than VQ-Rec on all datasets. The results indicate that VQ-Rec can learn and transfer general sequential patterns of discrete codes to downstream domains or platforms.

$(2)$ \emph{$w/o$ Semi-synthetic NS}: Removing semi-synthetic negative samples from the pre-training loss (Eqn.~\eqref{eqn:loss}), VQ-Rec may suffer from sparsity issues and get sub-optimal results. 

$(3)$ \emph{$w/o$ Fine-tuning}: The performance drops sharply if the pre-trained model is not fine-tuned, additionally showing that transferring to domains with varied semantics can be difficult.

$(4)$ \emph{Reuse PQ Index Set}: We directly encode the downstream item indices using the PQ centroids that are used for pre-training. The large semantic gap makes the indices follow a long-tail distribution. Due to the reduced distinguishability, this variant performs worse. 

$(5)$ \emph{$w/o$ Code-Emb Alignment}: In this variant, we remove matrices $\bm{\Pi}_k$ in Eqn.~\eqref{ekplus} that are used to align the pre-trained embeddings to downstream codes. The results show that permutation-based alignment network can generally improve performance.

$(6)$ \emph{Random Code}: In this variant, we randomly assign the pre-trained embeddings to downstream items. This variant generally has worse performance than the default method,
showing that the learned vector-quantized codes can preserve text characteristics. We also note that on Online Retail, this variant has slightly better performance, mainly because the item text is relatively short (\ie only $27.8$ words on average). The results suggest that informative item text is essential for deploying such text-based recommenders.

\subsection{Further Analysis}

\subsubsection{Capacity Analysis \wrt Increasing Training Data}

To show whether the proposed discrete code embeddings have a better capacity, we simulate a scenario where we have increasing training data. In detail, we train the models with different proportions of training interactions (\ie $20\%-100\%$) and present the performance on the test set in Figure~\ref{fig:exp-sparse}.
As we can see, the performance of VQ-Rec can always improve with more training data, outperforming the compared methods. The results indicate that VQ-Rec has a better capacity to fit training sequences.

\subsubsection{Transferability Analysis \wrt Downstream Datasets}

\begin{figure}[t]
	{
		\begin{minipage}[t]{0.8
		\linewidth}
			\centering
			\includegraphics[width=1\textwidth]{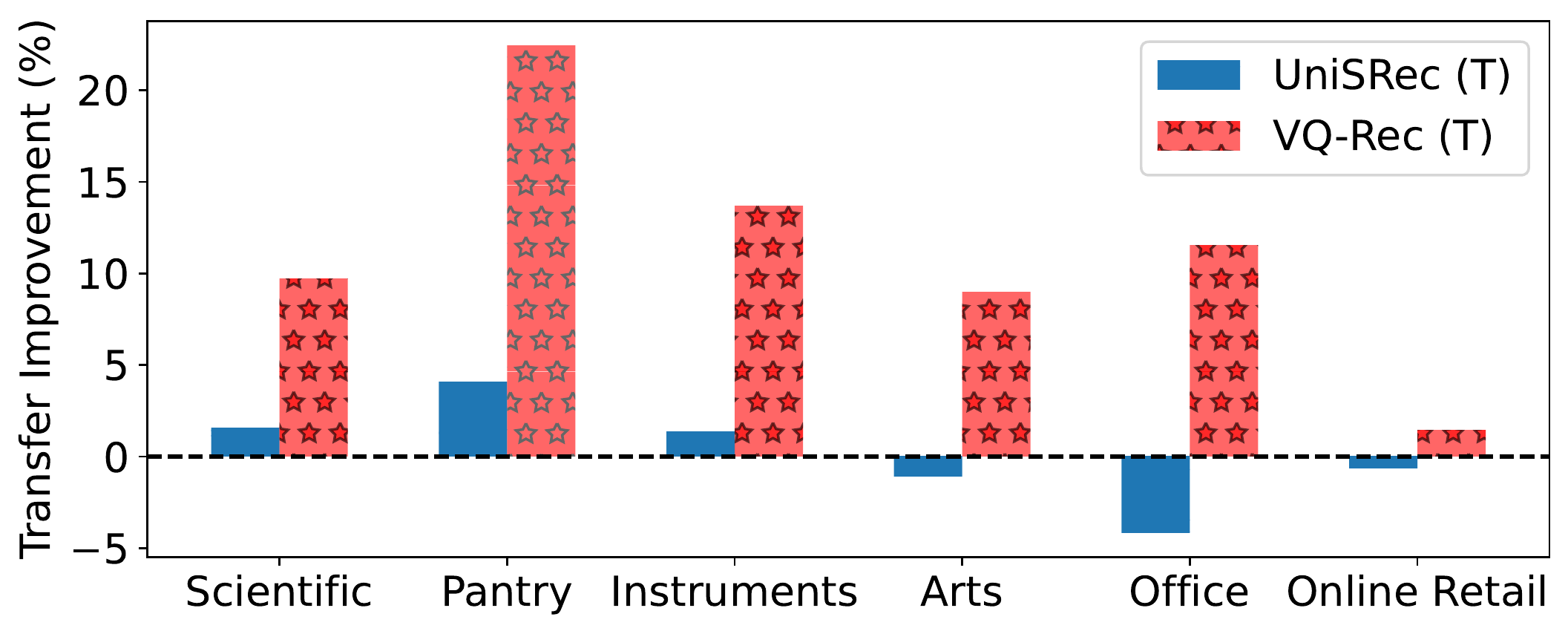}
		\end{minipage}
	}
	\caption{Relative improvement comparison ratios through transferring for Recall@10 \wrt different downstream datasets. Below the line denotes negative transfer.}
	\label{fig:exp-transfer}
\end{figure}

In this part, we show the relative transferring improvements (\ie $w/$ pre-training compared to $w/o$ pre-training) in Figure~\ref{fig:exp-transfer} on each downstream dataset. We can see that due to the capacity issue of PLM-based item representations, UniSRec may suffer from negative transfer issues on several datasets (\ie Arts, Office, and Online Retail). In contrast, VQ-Rec can benefit from pre-training on all six experimental datasets and has a maximum of 20+\% improvement. The results show that the proposed techniques can help recommenders transfer to those downstream scenarios with varied semantics.

\begin{figure}[t]
	{
		\begin{minipage}[t]{0.49\linewidth}
			\centering
			\includegraphics[width=1\textwidth]{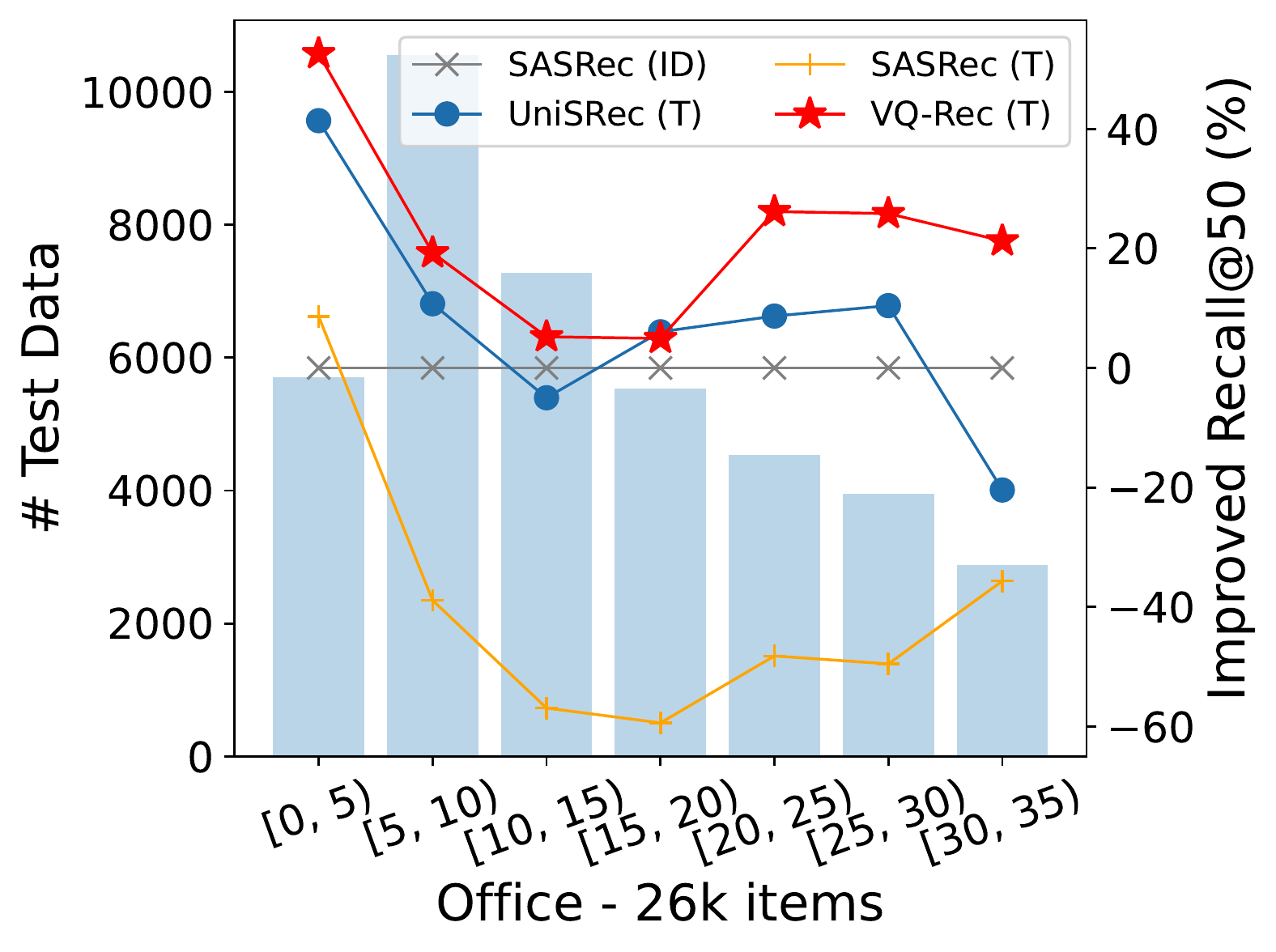}
		\end{minipage}
		\begin{minipage}[t]{0.49\linewidth}
			\centering
			\includegraphics[width=1\textwidth]{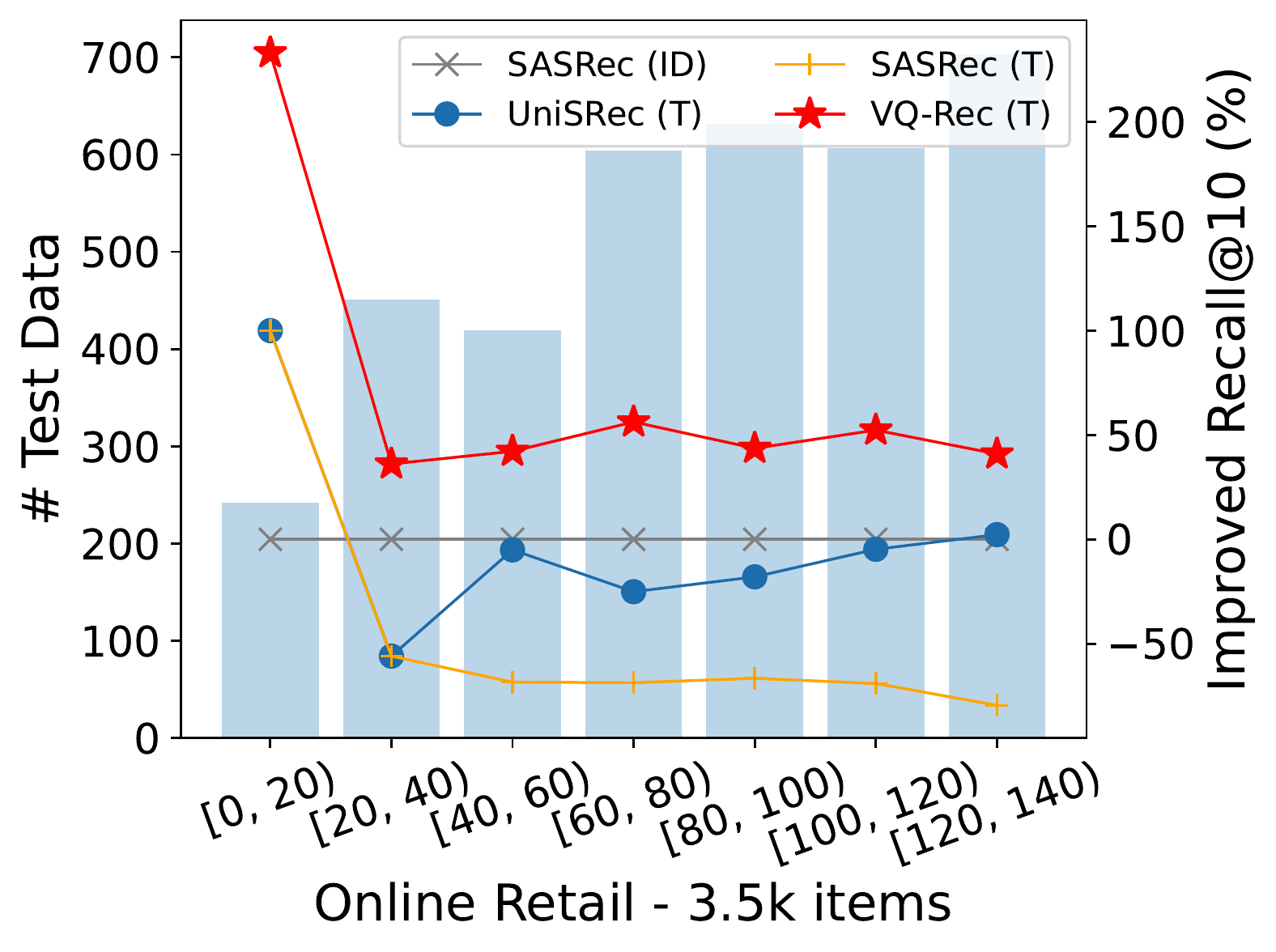}
		\end{minipage}
	}
	\caption{Performance comparison \wrt cold-start items on ``Office'' and ``Online Retail'' datasets. The bar graph denotes the number of interactions in test data for each group. The line chart denotes the relative improvement ratios compared with the baseline method SASRec (ID). 
	}
	\label{fig:exp-cold-start}
\end{figure}

\subsubsection{Cold-Start Item Analysis}

One of the motivations to develop transferable recommenders is to alleviate the cold-start recommendation issue. We split the test data into different groups according to the popularity of ground-truth items and present the results in Figure~\ref{fig:exp-cold-start}.
Although recommenders that directly map text to 
textual item representations have good performance on cold-start groups, \eg $[0,5)$ for Office and $[0, 20)$ for Online Retail, their performance drops on those popular groups.
In contrast, VQ-Rec has better performance than SASRec in all  groups, especially in cold-start groups. The results indicate that recommendations on these long-tail items may benefit 
from the proposed pre-training techniques. 

%% file: sec-related.tex
\section{Related Work}

\textbf{Sequential recommendation.} 
Sequential recommendation~\cite{rendle2010fpmc,hidasi2016gru4rec,kang2018sasrec} aims to predict 
the next interacted items based on historical interaction sequences.
Early works follow the Markov Chain 
assumption~\cite{rendle2010fpmc}, while recent studies mainly
focus on designing different neural network models, including Recurrent Neural Network (RNN)~\cite{hidasi2016gru4rec,li2017narm}, Convolutional Neural Network (CNN)~\cite{tang2018caser}, Transformer~\cite{kang2018sasrec,sun2019bert4rec,he2021locker,hou2022core}, Graph Neural Network (GNN)~\cite{wu2019srgnn,chang2021surge} and Multilayer Perceptron (MLP)~\cite{zhou2022fmlp}. However, most of these approaches are developed based on 
item IDs~\cite{kang2018sasrec} or attributes~\cite{zhang2019fdsa} defined on one specific domain, making it difficult to leverage behavior sequences from other domains or platforms. More recently, there have been attempts that employ textual or visual features as transferable item representations~\cite{ding2021zero,mu2022ida,hou2022unisrec,wang2022transrec}.
Furthermore, several studies propose constructing a unified model to solve multiple recommendation-oriented tasks 
based on PLMs~\cite{geng2022p5,cui2022m6}.
Our work is built on these studies, but has a different focus: we incorporate discrete codes to decouple the binding between text encodings and item representations, enhancing the representation capacity with specially designed pre-training and fine-tuning strategies.

 \textbf{Transfer learning for recommendation.} To alleviate the data sparsity and cold-start issues that broadly exist in recommender systems, researchers have explored the idea of knowledge transfer from other domains~\cite{zhu2019dtcdr,zhu2021crossdomain,xie2021ccdr}, markets~\cite{bonab2021crossmarket} or platforms~\cite{lin2019crossplatform}.
 Existing approaches mainly rely on shared information between the source and target domains to conduct the transfer, \eg common users~\cite{hu2018conet,yuan2021one,xiao2021uprec,wu2020ptum}, items~\cite{singh2008cmf,zhu2019dtcdr} or attributes~\cite{tang2012cdcr}. Recently, pre-trained language models~\cite{devlin2019bert,liu2019roberta} (PLM) have shown to be general semantic bridges to connect different tasks or domains~\cite{brown2020gpt3}, and several studies propose to encode the associated text of items by PLMs as universal item representations~\cite{ding2021zero,hou2022unisrec,geng2022p5}. Based on pre-training universal item representations, one can 
 transfer the fused knowledge to downstream domains without overlapping users or items. 
 However, these studies usually enforce a tight binding between the text encodings from PLMs and the final item representations, which might over-emphasize the effect of text features. In contrast, we propose a novel two-step representation scheme based on discrete codes, which is endowed with a more strong representation capacity for enhancing the cross-domain recommendation. %

 \textbf{Sparse representation for recommendation.} Learning sparse codes~\cite{SL-survey1,SL-survey2,SL-survey3,SL-survey4} is a widely adopted way to represent data objects in machine learning, inspiring a variety of studies related to product quantization~\cite{jegou2010pq,ge2013opq,PQ2,PQ3}, multi-way compact embedding~\cite{CompactEmbedding,chen2020dpq}, semantic hashing~\cite{SemanticHashing,kitaev2020reformer}, etc. In contrast to continuous representation, it seeks a sparse scheme for capturing  the most salient representation dimensions. Specifically, discrete representations are also applied in recommender systems~\cite{DC-rec1,DC-rec2,DC-rec3,DC-rec4,DC-rec5,DC-rec6,wu2021linear}, and existing studies mainly aim to develop efficient recommendation algorithms in both \emph{memory} and \emph{time} based on the sparse representations, so as to develop large-scale recommender systems~\cite{DC-large0,DC-large1,DC-large2,DC-large3}. Different from these studies, our work aims to leverage the generality of text semantics for learning transferable item representations based on a pre-trained approach.